\documentclass{iopart}
\usepackage{graphicx} 
\usepackage{graphics} 
\usepackage{amssymb,amsfonts,latexsym}
\usepackage{dcolumn,epsfig}
\usepackage{setspace}
\usepackage{subfigure}
\usepackage{url}
\usepackage{citesort}
\usepackage{rotating}
\usepackage[usenames]{color}

\usepackage{ulem}
\newcommand{\beq}{\begin{equation}}
\newcommand{\eeq}{\end{equation}}
\newcommand{\bea}{\begin{eqnarray}}
\newcommand{\eea}{\end{eqnarray}}

\newcommand{\Ytwo}{{{}^{-2}Y}}

\def\IL{\relax{\rm I\kern-.18em L}}

\newcommand{\be}{\begin{equation}}
\newcommand{\ee}{\end{equation}}
\newcommand{\bel}[1]{\begin{equation}\label{#1}}
\newcommand{\ba}{\begin{eqnarray}}
\newcommand{\ea}{\end{eqnarray}}
\newcommand{\bal}[1]{\begin{eqnarray}\label{#1}}

\newcommand{\Ms}{M_{\odot}}

\def\ltsima{$\; \buildrel < \over \sim \;$}
\def\simlt{\lower.5ex\hbox{\ltsima}}
\def\gtsima{$\; \buildrel > \over \sim \;$}
\def\simgt{\lower.5ex\hbox{\gtsima}}

\begin{document}

\title[NINJA]{Status of NINJA: the Numerical INJection Analysis project}
\author{
Benjamin~Aylott${}^{1}$,
John~G.~Baker${}^{2}$,
William~D.~Boggs${}^{3}$,
Michael~Boyle${}^{4}$,
Patrick~R.~Brady${}^{5}$,
Duncan~A.~Brown${}^{6}$,
Bernd~Br\"ugmann${}^{7}$,
Luisa~T.~Buchman${}^{4}$,
Alessandra~Buonanno${}^{3}$,
Laura~Cadonati${}^{8}$,
Jordan~Camp${}^{2}$,
Manuela~Campanelli${}^{9}$,
Joan~Centrella${}^{2}$,
Shourov~Chatterji${}^{10,11}$,
Nelson~Christensen${}^{12}$,
Tony~Chu${}^{4}$,
Peter Diener${}^{13,14}$,
Nils~Dorband${}^{15}$,
Zachariah~B.~Etienne${}^{16}$,
Joshua~Faber${}^{9}$,
Stephen~Fairhurst${}^{17}$,
Benjamin~Farr${}^{9,17}$,
Sebastian~Fischetti${}^{8}$,
Gianluca~Guidi${}^{10,18}$,
Lisa~M.~Goggin${}^{5}$,
Mark~Hannam${}^{19}$,
Frank~Herrmann${}^{20,30}$,
Ian~Hinder${}^{20}$,
Sascha~Husa${}^{21,15}$,
Vicky~Kalogera${}^{22}$,
Drew~Keppel${}^{11}$,
Lawrence~E.~Kidder${}^{23}$,
Bernard~J.~Kelly${}^{2}$,
Badri~Krishnan${}^{15}$,
Pablo~Laguna${}^{24}$,
Carlos~O.~Lousto${}^{9}$,
Ilya~Mandel${}^{22}$,
Pedro~Marronetti${}^{25}$,
Richard~Matzner${}^{29}$,
Sean~T.~McWilliams${}^{2}$,
Keith~D.~Matthews${}^{4}$,
R.~Adam~Mercer${}^{5}$,
Satyanarayan~R.~P.~Mohapatra${}^{8}$,
Abdul H. Mrou{\'e}${}^{23}$,
Hiroyuki~Nakano${}^{9}$,
Evan~Ochsner${}^{3}$,
Yi~Pan${}^{3}$,
Larne~Pekowsky${}^{6}$,
Harald~P.~Pfeiffer${}^{4}$,
Denis~Pollney${}^{15}$,
Frans~Pretorius${}^{26}$,
Vivien~Raymond${}^{22}$,
Christian~Reisswig${}^{15}$,
Luciano~Rezzolla${}^{15}$,
Oliver~Rinne${}^{27}$,
Craig~Robinson${}^{12}$,
Christian~R\"over${}^{28}$,
Luc{\'i}a~Santamar{\'i}a${}^{15}$,
Bangalore~Sathyaprakash${}^{17}$,
Mark~A.~Scheel${}^{4}$,
Erik~Schnetter${}^{13,14}$,
Jennifer~Seiler${}^{15}$,
Stuart~L.~Shapiro${}^{16}$,
Deirdre~Shoemaker${}^{24}$,
Ulrich~Sperhake${}^{4,7}$,
Alexander~Stroeer${}^{31,2}$,
Riccardo~Sturani${}^{10,18}$,
Wolfgang~Tichy${}^{25}$,
Yuk~Tung~Liu${}^{16}$,
Marc~van~der~Sluys${}^{22}$,
James~R.~van Meter${}^{2}$,
Ruslan~Vaulin${}^{5}$,
Alberto~Vecchio${}^{1}$,
John~Veitch${}^{1}$,
Andrea~Vicer\'e${}^{10,18}$,
John~T.~Whelan${}^{9,15}$,
Yosef~Zlochower${}^{9}$
} 

\address{$^{1}$ School of Physics and Astronomy, University of Birmingham, Edgbaston, Birmingham B15 2TT, UK}
\address{$^{2}$ NASA Goddard Space Flight Center, Greenbelt, MD 20771, USA}
\address{$^{3}$ Maryland Center for Fundamental Physics, Department of Physics,
University of Maryland, College Park, MD 20742, USA}
\address{$^{4}$ Theoretical Astrophysics 130-33, California Institute of Technology, Pasadena, CA 91125}
\address{$^{5}$ University of Wisconsin-Milwaukee, P.O.~Box 413, Milwaukee, WI 53201, USA}
\address{$^{6}$ Department of Physics, Syracuse University, Syracuse, New York, 13254}
\address{$^{7}$ Theoretisch Physikalisches Institut, Friedrich Schiller Universit\"at, 07743 Jena, Germany}
\address{$^{8}$ Department of Physics, University of Massachusetts, Amherst, MA 01003}
\address{$^{9}$ Center for Computational Relativity and Gravitation and School of Mathematical Sciences, Rochester Institute of Technology, 85 Lomb Memorial Drive, Rochester, NY 14623}
\address{$^{10}$ INFN-Sezione Firenze/Urbino, I-50019 Sesto Fiorentino, Italy}
\address{$^{11}$ LIGO -- California Institute of Technology, Pasadena, CA 91125, USA}
\address{$^{12}$ Physics \& Astronomy, Carleton College, Northfield, MN, USA}
\address{$^{13}$ Center for Computation \& Technology, Louisiana State University, Baton Rouge, LA 70803}
\address{$^{14}$ Department of Physics \& Astronomy, Louisiana State University, Baton Rouge, LA 70803}
\address{$^{15}$ Max-Planck-Institut f\"ur Gravitationsphysik, Am M\"uhlenberg 1, D-14476 Potsdam, Germany.}
\address{$^{16}$ Department of Physics, University of Illinois at Urbana-Champaign, Urbana, IL 61801}
\address{$^{17}$ School of Physics and Astronomy, Cardiff University, The Parade, Cardiff, UK}
\address{$^{18}$ Istituto di Fisica, Universit\`a di Urbino, I-61029 Urbino, Italy}
\address{$^{19}$ Physics Department, University College Cork, Cork, Ireland}
\address{$^{20}$ Center for Gravitational Wave Physics, The Pennsylvania State University, University Park, PA 16802}
\address{$^{21}$ Departament de F\'isica, Universitat de les Illes Balears, Palma de Mallorca, Spain}
\address{$^{22}$ Department of Physics and Astronomy, Northwestern University, Evanston, IL, USA}
\address{$^{23}$ Center for Radiophysics and Space Research, Cornell University, Ithaca, New York, 14853}
\address{$^{24}$ Center for Relativistic Astrophysics and School of Physics, Georgia Institute of Technology, Atlanta, GA 30332}
\address{$^{25}$ Department of Physics, Florida Atlantic University, Boca Raton, FL 33431}
\address{$^{26}$ Department of Physics, Princeton University, Princeton, NJ 08540}
\address{$^{27}$ Department of Applied Mathematics and Theoretical Physics, Centre for Mathematical Sciences, Wilberforce Road, Cambridge CB3 0WA, UK, and King's College, Cambridge CB2 1ST, UK}
\address{$^{28}$ Max-Planck-Institut f\"ur Gavitationsphysik, Hannover, Germany}
\address{$^{29}$ University of Texas at Austun, Austin, Texas, 78712}
\address{$^{30}$ Center for Scientific Computation and Mathematical Modeling, University of Maryland, 4121 CSIC Bldg. 406, College Park, Maryland 20742, USA}
\address{$^{31}$ CRESST, University of Maryland, College Park, Maryland 20742, USA}

\begin{abstract}
The 2008 NRDA conference introduced the Numerical INJection Analysis project (NINJA), a new collaborative effort between the numerical relativity community and the data analysis community. NINJA focuses on modeling and searching for gravitational wave signatures from the coalescence of binary system of compact objects.  
We review the scope of this collaboration and the components of the first NINJA project, where numerical relativity groups shared waveforms and data analysis teams applied various techniques to detect them when embedded in colored Gaussian noise.  
\end{abstract}

\section{Introduction}\label{s:intro}
The coalescence of binary systems of compact objects is one of the most promising sources of gravitational wave radiation for ground-based detectors.  
The past few years have witnessed the successful construction and operation of a world-wide network of large interferometric gravitational wave detectors, with the three LIGO detectors in the United States~\cite{Abbott:2007kva}, Virgo in Italy~\cite{VIRGOstatus}, TAMA in Japan~\cite{Tsubono:2000if} and GEO600 in Germany~\cite{Willke:2007zz}. 
A number of searches for unmodelled bursts and binary coalescences are in
progress and several results have
already been reported~\cite{Abbott:2003pj,Abbott:2005pe,Abbott:2005pf,%
Abbott:2007xi,Abbott:2007ai,Abbott:2009tt,Abbott:2004rt,Abbott:2005fb,Abbott:2005at,%
Abbott:2005rt,Abbott:2005qm,Abbott:2007wu,Ando:2004rr,Ando:2004rr,Akutsu:2006ei,Acernese:2007zz,Acernese:2007ek,Acernese:2008cs,Acernese:2008dd}.
In parallel, recent breakthroughs in numerical relativity made it possible to   successfully simulate the merger phase of Binary Black Hole (BBH) coalescences.  Following the pioneer work of Pretorius \cite{Pretorius:2005gq}, and
the independent results from the Goddard~\cite{Baker:2005vv} and Brownsville (now at RIT)~\cite{Campanelli:2005dd} groups, several
numerical relativity groups around the world have successfully evolved
various BBH configurations.
This has lead to important new physical insights, 
such as the prediction of large recoil velocities
due to asymmetric emission of gravitational radiation during the
merger~\cite{Herrmann:2006ks,Baker:2006vn,Herrmann:2007ac,Koppitz:2007ev,Campanelli:2007ew,Gonzalez:2007hi,Campanelli:2007cga,Baker:2007gi,Herrmann:2007ex,Brugmann:2007zj,Schnittman:2007ij,Pollney:2007ss,Lousto:2007db,Baker:2008md,Dain:2008ck,Healy:2008js,Gonzalez:2008bi}
and the prediction of the parameters of the remnant for a wide class of initial states~\cite{Campanelli:2006fg,Gonzalez:2006md,Campanelli:2006fy,Rezzolla:2007xa,Boyle:2007sz,Rezzolla:2007rd,Marronetti:2007wz,Rezzolla:2007rd,Sperhake:2007gu,Hinder:2007qu,Boyle:2007ru,Tichy:2008du,Rezzolla:2008sd}. 
A review of the current status of BBH simulations is available in this volume~\cite{Hannam:2009rd}. 

Numerical relativity now provides a complete model of the coalescence waveform, inclusive of the merger.  
However, the numerical relativity results have not yet been synthesized into an analytical model over a broad range of mass ratios, spin, and eccentricity.
Furthermore, despite significant progress, there is not yet a complete description of how post-Newtonian and numerical simulations are to be matched with each other, over the full parameter space.  
Most searches for gravitational waves from BBH mergers have so far relied on the post-Newtonian analytical description of the expected gravitational wave signal, valid only when the black holes are sufficiently far apart. 
A verification of their robustness is urgent, and,
more generally, it is important to quantify the performance of the data analysis pipelines for detection and parameter estimation, to set astrophysical upper limits, and to aid in follow-up studies and the interpretation of potential detection candidates.

The Numerical INJection Analysis project (NINJA), open to all scientists interested in numerical simulations and gravitational wave data analysis, was started in the spring of 2008 to address these issues and form a new, close collaboration between the two communities. 
The first NINJA project benefited from the contribution of 10 numerical relativity groups and 9 data analysis teams, for a total of 76 participants from 30 institutions.
In a step towards the incorporation of numerical relativity waveforms in gravitational wave data analysis, various data analysis algorithms analyzed BBH coalescence waveforms buried in simulated Gaussian noise at the design sensitivity of initial LIGO and Virgo. 
The first NINJA project only includes BBH simulations, but it is expected that future NINJA analyses will be expanded to include, for example, binary neutron star and supernovae simulations.

The first NINJA project was a learning experience which brought the two communities closer. To facilitate this, minimal constraints were imposed: each numerical relativity group chose which waveforms to share and each data analysis team chose which methods and results to contribute. We learned about technical and conceptual issues, and how to address them. However, due to the small statistics of simulations and a lack of systematic studies, comparisons and conclusions drawn from the first NINJA project are limited and should be handled with care,  as in most cases they are only the first steps towards fully understanding the sensitivity of data-analysis pipelines to black hole signals.

This proceedings paper provides a highlight overview of the NINJA collaboration and its first project. For details on the waveforms, on the data analysis techniques and the results, we refer to~\cite{Aylott:2009ya}.

\section{Numerical Relativity Waveforms}

Ten numerical relativity teams contributed BBH coalescence waveforms that are solutions to Einstein's equations, with no restrictions on their morphology or accuracy.  
Each team submitted a maximum of two waveforms, or up to five waveforms if they were part of a one-parameter family, following the format specifications in~\cite{Brown:2007jx}. 

The NINJA waveforms cover a variety of physical and numerical parameters. 
Most simulations model low-eccentricity inspiral, with mass ratio $q = M_1/M_2$ from 1 to 4, and several spin configurations;  
the initial frequency of the $\ell=m=2$ mode ranges from $0.033/M$ to $0.203/M$, where $M$ is the 
sum of the initial black-hole masses, and the waveform length varies between a few 100M and over 4000M. 
The different contributions are:
BAM\_HHB~\cite{Brugmann:2008zz,Husa:2007hp,Hannam:2007ik,Hannam:2007wf,Bruegmann:2003aw} 
and BAM\_FAU~\cite{Brugmann:2008zz,Husa:2007hp,Tichy:2008du,Bruegmann:2003aw}, using the {\tt BAM} code,
the AEI/LSU code {\tt CCATIE}~\cite{Alcubierre:2000xu,Alcubierre:2002kk,Koppitz:2007ev,Pollney:2007ss,Rezzolla:2007xa},
the Goddard Space Flight Center's code {\tt Hahndol}~\cite{Imbiriba:2004tp,vanMeter:2006vi}, 
the RIT code {\tt LazEv}~\cite{Zlochower:2005bj,Campanelli:2005dd,Dain:2008ck}, 
Ulrich Sperhake's code {\tt Lean}~\cite{Sperhake:2006cy,Sperhake:2007gu,Sperhake:2008ga}, 
the Georgia Tech/Penn State code {\tt MayaKranc}~\cite{Vaishnav:2007nm,Hinder:2007qu}, 
the Princeton University code~\cite{Pretorius:2004jg,Pretorius:2005gq,Buonanno:2006ui,Pretorius:2007jn},
the Cornell/Caltech collaboration 
code {\tt SpEC}~\cite{Scheel:2006gg,Pfeiffer:2007yz,Boyle:2007ft,Scheel:2008rj},
and the University of Illinois at
Urbana-Champaign code~\cite{Etienne:2007hr}.

The numerical codes follow one of two approaches to solving the Einstein equations:  the generalized harmonic formulation~\cite{Pretorius:2005gq}, or  the moving-puncture approach~\cite{Campanelli:2005dd,Baker:2005vv}.  
All of the results make the simplifying assumption of conformal flatness for the spatial metric of the initial slice, which leads to some spurious gravitational radiation in the initial data, and attempt to model non-eccentric inspirals, with the exception of PU--T52W and {\tt MayaKranc}--e02 that target eccentric inspirals.  
To estimate the gravitational-wave signal at a finite distance from the source, the {\tt SpEC} and {\tt CCATIE} contributions use the Zerilli-Moncrief perturbative formalism~\cite{Moncrief:1974am,Nagar:2005ea,Sarbach:2001qq}, while all others use the Newman-Penrose curvature scalar $\psi_4$~\cite{baker-2002-65}. 
Details on the implementations within particular codes can be found, for instance in refs.~\cite{Baker:2001sf,Sperhake:2006cy,Pollney:2007ss,Brugmann:2008zz}; a detailed review of similarities and differences between NINJA waveforms, codes and numerical methods, is available in ref.~\cite{Aylott:2009ya}.

Waveforms have been contributed in the form of spherical harmonic modes $\Ytwo_{lm}$ of spin-weight $-2$ of the radiation field at large distance from the source, as specified in~\cite{Brown:2007jx}.  
In the Transverse-Traceless (TT) gauge, the spatial components $h_{ij}$ are:
\begin{equation}
  h_{ij} = A_{ij}\frac{M}{r} + \mathcal{O}\left(r^{-2}\right)\,,
\end{equation}
where $M$ is the total mass of the system, $r$ is the distance from the source, and $A_{ij}$ is a time dependent TT tensor.  In the TT gauge, $h_{ij}$ has two independent polarizations $h_+$ and
$h_\times$, and the complex function $h_+-ih_\times$ is decomposed as:
\begin{equation}
\label{eq:mode-decomposition}
  h_+ - ih_\times = \frac{M}{r}\sum_{\ell=2}^{\infty}\sum_{m=-\ell}^\ell H_{\ell m}(t)\,
  \Ytwo_{\ell m}(\iota,\phi)\,.
\end{equation}
The expansion parameters $H_{lm}$ are complex functions of the retarded time $t-r$ and, if $r$ is the radius of extraction of the wave,  $H_{lm}$ is function of $t$ only. 
The angles $\iota$ and $\phi$ are respectively the polar and azimuthal angles in a suitable coordinate system centered on the source.  
Computation of the strain from the Zerilli-Moncrief odd- and even-parity ($Q^\times_{lm}$, $Q^+_{lm}$) multipoles of the metric perturbation requires one time integration~\cite{Nagar:2005ea,Pollney:2007ss}; computation of the strain from the Newman-Penrose curvature scalar $\psi_4$ requires two time integrations, with a proper choice of the constants of integration, and attention to artifacts resulting from the finite extraction radii.

While no attempt was made, in this first project, to restrict waveform parameters, future exchanges will be coordinated to address questions such as how many modes are needed for complicated waveforms: should a fixed $\ell$ cutoff be imposed or  should all modes with a minimum percentage fraction of the total energy be used? What is the impact of this on the total systematics as a function of mass and the detector noise curve? 
We will also explore hybridization with post-Newtonian waveforms, to avoid abrupt startup of the waveforms, and exploit overlaps and complementarities between teams and codes, for a better coverage of the parameter space.

\section{The NINJA Data Set}\label{s:data}

Following the format outlined in~\cite{Brown:2007jx}, numerical relativity groups provided one ASCII data file for each mode $(\ell,m)$ with appreciable contribution to the waveform. Each data file consists of three columns: time in units of the total
mass, and the real and imaginary parts of the mode coefficients $H_{\ell m}$.

The NINJA data set simulated signals as seen by the gravitational wave detectors, which also depend on the total mass of the binary system, its  distance, and its orientation with respect to a particular detector.  
The data production starts from a Monte Carlo population of binary black hole systems, with the source distance logarithmically distributed in $50-500\,$Mpc. Orientations are uniformly distributed and masses are uniform in $20 - 350M_\odot$. Since the total mass $M$ scales both time and amplitude, each waveform can be scaled to an arbitrary value of total mass, with the requirement that the initial frequency  of the dominant $\ell=m=2$ mode is below 30~Hz, to avoid startup transient artifacts. This is an appropriate threshold for the sensitivity curve of the LIGO and Virgo initial detectors, and sets a lower limit on the value of $M$
for each waveform: the longer the waveform, the lower the value of $M$ it can be injected at.  
No restrictions were placed on other simulation parameters, such as spin, mass-ratio and eccentricity, as these were determined by the numerical waveform parameters.

Given $H_{\ell m}(t)$, the total mass, the distance to the source, the angles $(\iota,\phi)$, 
$h_{+,\times}$ (from eq.~\ref{eq:mode-decomposition}), and the detector response functions $F_{+,\times}$ the observed strain is:
\begin{equation}
  h(t) = h_+(t) F_+(\Theta,\Phi,\psi) + h_\times(t) F_\times(\Theta,\Phi,\psi)\, ,
\end{equation}
where $(\Theta,\Phi)$ are sky angles in the detector frame and $\psi$ is the polarization angle. The time $t$ is in units of seconds.  

\begin{figure}
  \centering
  \includegraphics[width=\textwidth]{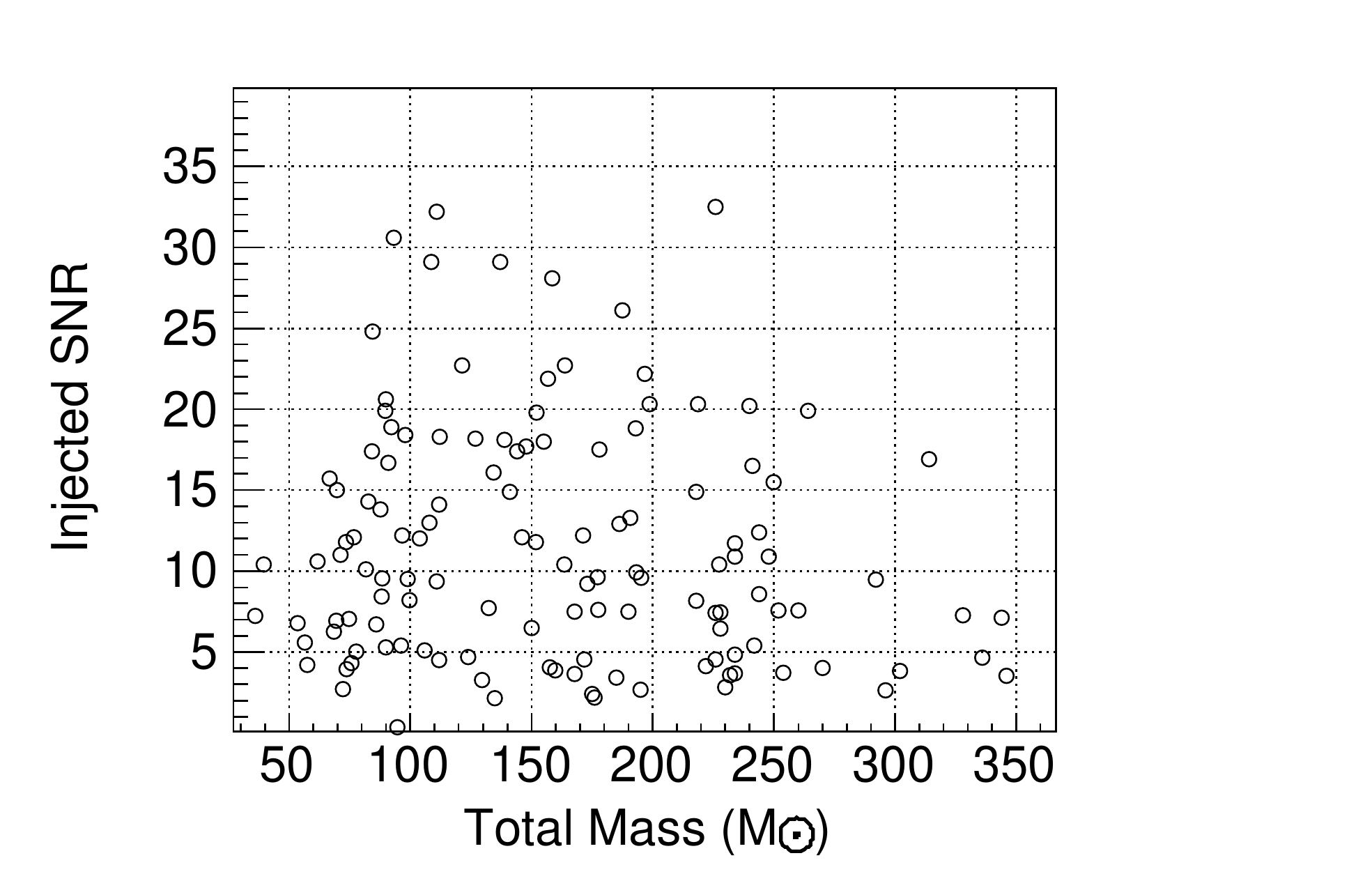}
  \caption{Distribution of signal-to-noise ratio versus total mass for the NINJA data set. 
  The SNR is detector-dependent; in this plot, it is computed at the LIGO-Hanford 4km detector. }
  \label{fig:injscat}
\end{figure}

Once signals for the chosen populations are produced as time series at $4096\,$Hz sampling rate, typical for LIGO and Virgo searches, they are added to stationary Gaussian noise colored with the initial LIGO and Virgo design power spectral densities. A data set was produced for each of the three LIGO interferometers and for the Virgo interferometer, with approximately 12 signals from each contributing group. 
The resulting NINJA data set consisted of 126 signals injected in a little over $30\,$hours; the time interval betwen adjacent signals is a random number between $700\pm 100\,$sec.  
The distribution of signal-to-noise ratios is detector-dependent; Fig.~\ref{fig:injscat} shows  SNR versus total mass at the LIGO Hanford 4km detector.

\section{Data Analysis}\label{s:analysis}

Nine data analysis teams contributed a variety of techniques to the NINJA project. 
Participants were provided with the data set described in section~\ref{s:data} and a list of  injected signals parameters, but did not have access to the raw waveforms.
The two fundamental data analysis goals are detection and parameter estimation. 
Detection has been pursued in the NINJA project with both matched-filtering and un-modeled, excess-power techniques.  Bayesian model and parameter estimation techniques were also applied to the NINJA data set.   
%
%
In this section we provide an overview of the techniques adopted; for more details and results, see~\cite{Aylott:2009ya}.
%

\subsection{Matched filter, modeled searches}

The LIGO Scientific Collaboration (LSC) and the Virgo Collaboration have implemented a matched filter analysis pipeline in previous searches for gravitational wave signatures from compact binary coalescences~\cite{LIGOS3S4Tuning,Abbott:2007xi}. 
For each waveform model, a \emph{bank} of templates covers the parameter space so that the fractional loss in signal-to-noise ratio (SNR) between any signal and the nearest template is less than $3\%$.
Data from each detector is match filtered against this bank~\cite{Allen:2005fk,Creighton:1999pm} and a \emph{trigger} is produced whenever the SNR exceeds a threshold of $5.5$. 
Triggers which do not have coincident parameters in two or more detectors are discarded~\cite{Robinson:2008un,Goggin:2008}; the remaining triggers are ranked by the square-sum of the SNRs of the triggers in the coincidence.  
For inspiral matched filter analyses, coincident inspiral triggers are subject to a second filtering stage, in which signal-based vetoes are also applied to separate true signals from noise fluctuations~\cite{Allen:2004gu,Rodriguez:2007}.  
A threshold is applied to an effective SNR $\rho_\mathrm{eff}$, which combines the matched filter SNR and the value of the $\chi^{2}$ signal-based veto~\cite{Allen:2004gu}.   

Six groups contributed matched filter searches to the NINJA project; the results can be divided into three categories based on the waveform templates used. 
Within these categories, different parameter choices were made to explore how the pipeline can detect numerical relativity simulations.

\subsubsection{Inspiral only waveforms from the Post-Newtonian expansion.}

The standard template for the LSC-Virgo search~\cite{Abbott:2003pj,Abbott:2005pe,Abbott:2005pf,Abbott:2007xi,Abbott:2009tt} is based on the non-spinning post-Newtonian inspiral waveforms, calculated directly in the Fourier domain through the stationary phase approximation~\cite{Droz:1999qx,Allen:2005fk}. 
These waveforms, referred to as SPA or TaylorF2, are parameterized by the binary's component masses $m_1$ and $m_2$~\cite{Owen:1998dk,Babak:2006ty,Cokelaer:2007kx}, or equivalently by total mass $M =m_1 + m_2$ and symmetric mass ratio $\eta = m_1 m_2 / M^2$. The amplitude evolution is modelled to leading order and the phase evolution is modelled to a specified post-Newtonian order.
Formally, the TaylorF2 waveform can be extended to arbitrary frequencies.  However, at higher frequencies, the post-Newtonian expansion does not accurately model the physics and the waveforms are terminated at a cutoff frequency $f_\mathrm{c}$. In the LSC-Virgo analyses, this is chosen to be the innermost stable circular orbit (ISCO) frequency for a test mass in a Schwarzschild spacetime.
Details of a search of the NINJA data using TaylorF2 waveforms are available in this volume \cite{Farr:2009pg}.

\subsubsection{Ringdown-only waveforms from black hole perturbation theory.} 
Ringdown searches use a two-parameter template bank parameterized by frequency and quality factor, constructed to cover the desired range of mass and spin~\cite{Creighton:1999pm}. 
The LSC ringdown matched filter algorithm~\cite{Goggin:2008} was used for the NINJA analysis, using a bank with frequency between 50 Hz and 2 kHz and quality factor between 2 and 20.

Assuming that the dominant ringdown mode is the fundamental mode, $\ell =m=2$, the frequency and quality factor of the waveform can be expressed in terms of the mass $M$ and dimensionless spin factor $a$ of the black hole using Echeverria's fit \cite{Echeverria:1989hg} to Leaver's numerical calculations \cite{Leaver:1985ax}:
\begin{eqnarray}
f&=&\frac{1}{2 \pi} \frac{c^3}{GM}\left[ 1-0.63\left( 1-a\right)
^{\frac{3}{10}}\right]  \label{eqn:Echeverria_fofMa} \\
Q&=&2\left( 1-a\right) ^{-\frac{9}{20}}.
\label{eqn:Echeverria_Qofa}
\end{eqnarray}
These equations can be inverted to calculate the mass $M$ and spin $a$ from the template parameters of a given trigger.

\subsubsection{Phenomenological/analytical models for the full coalescence.}

Post-Newtonian theory is valid when the black holes are sufficiently separated, but becomes unreliable as the velocity of the black holes increase in their final orbits before merger and the non-perturbative information contained in numerical simulations becomes important.  A successful approach has been to combine analytical (PN) and numerical (NR) results into full waveform templates.  Three different families of waveforms have been used in analyzing the NINJA data: Extended TaylorF2, EOBNR and phenomenological waveforms.    

Recent comparisons to numerical relativity waveforms have shown that extending the TaylorF2 waveforms to higher frequencies improves their sensitivity at higher masses~\cite{Pan:2007nw,Boyle:2009dg}.  
An analysis using these extended waveforms was performed on the NINJA data.
Specifically, the cutoff frequency $f_\mathrm{c}$ was increased from ISCO to the effective ringdown (ERD) frequency, obtained by comparing post-Newtonian models to numerical waveforms \cite{Pan:2007nw}.  This increases the sensitivity of the search as the extended templates can detect some of the power contained in the late inspiral or early merger part of the signal. 

Two studies of the match between numerical waveforms~\cite{Pan:2007nw,Boyle:2009dg} and TaylorF2 templates suggest that the search efficiency can be improved by extending the range of $\eta$ outside the physical range of $\eta \le 0.25$, and terminating the waveforms at a weighted ringdown (WRD) frequency, between ISCO and ERD \cite{Boyle:2009dg}.  By extending the template bank to cover all points with $\eta \le 1$, its size was approximately doubled.  An analysis of the NINJA data with the extended template bank showed an increse in the recovered SNR for some signals.

Combining results from PN and perturbation theory, the EOB
model~\cite{Buonanno:1998gg,Buonanno:2000jz} predicted inspiral, merger and
ringdown waveforms.  The non-spinning EOB model has been further improved by
calibrating it to NR results, achieving rather high matching perfomances
without maximizing on binary parameters, but only on initial phase and time of
arrival~\cite{Buonanno:2006ui,Pan:2007nw,Buonanno:2007pf,Damour:2007yf,Damour:2007vq,Damour:2008te,Boyle:2008ge}.
These waveforms, called EOBNR, were also used in analyzing the NINJA data. 

The phenomenological approach~\cite{Ajith:2007kx,Ajith:2007qp} matches PN and NR waveforms in a regime where both are sufficiently accurate, and fits the resulting  \emph{hybrid} waveform to a phenomenological model in frequency domain. This procedure has been carried out for non-spinning black holes; each waveform is parametrized by the physical parameters of the system, i.e. the masses $m_1$ and $m_2$ of the black holes.  Results of a search of the NINJA data with phenomenological templates are given in~\cite{Santamaria:2009tm}.

\subsection{Unmodeled searches}

{\it Burst} searches are designed to detect gravitational wave transient signatures with minimal assumptions on their origin and waveform. They do not use templates and instead target excesses of power in the time-frequency plane. %
Since they do not assume a template, and they target short transients, burst searches are suited for the detection of the merger phase of the coalescence, and have the potential to probe a large parameter space, inclusive of spin and ellipticity, at no additional computational cost.
Two such algorithms, developed by the LSC and Virgo, analyzed the NINJA data: Q-pipeline and HHT.

The Q-pipeline~\cite{Chatterji:2005,Chatterji:2004qg,LSCburstS5y1} is a multi-resolution time-frequency search for statistically significant excess signal energy, equivalent to a templated matched filter search for sinusoidal Gaussians in whitened data. The template
bank covers a finite region in time, frequency, and quality factor such that mismatch between any sinusoidal Gaussian in this signal space and the nearest basis function does not exceed a
maximum of 20\% in energy. For the purpose of the NINJA project, the Q-pipeline analysis focused on the detection efficiency at the single detector, with an SNR threshold of 5.5, as in the matched filter searches.

The Hilbert-Huang Transform (HHT)~\cite{Huang:1998,Camp:2007ee,Stroeer:2008} is an adaptive algorithm that decomposes the data into Intrinsic Mode Functions (IMF), each representing a unique locally monochromatic frequency scale of the data. The original data is recovered with a sum over all IMFs.
The Hilbert transform, applied to each IMF, provides instantaneous frequency and amplitude as a function of time, thus providing high time-frequency resolution for signal detection.
Details on how this pipeline was tested on NINJA data are available in this volume, in ref.~\cite{hht:astroeer2008}.

\subsection{Comparison of searches}
The analysis techniques applied to NINJA data have comparable performance, if the same SNR threshold is used.   
However, the simulation statistics in this first NINJA project were quite limited, and the SNR of injected signals was relatively large, as shown in Fig.~\ref{fig:injscat}. 
Consequently, no strong conclusions can be drawn from these results except that unmodeled algorithms and matched filtering to ringdown or to the ``wrong'' templates can still identify loud signals in Gaussian noise, although not necessarily with the correct parameters, as they  preferentially detect whichever portion of the signal is in the detectors' most sensitive band. For the lowest injected masses, this corresponds to the inspiral phase and the detected frequency is close to $f_{ISCO}$, but as the injected mass increases, the ringdown dominates, as expected. 
For a quantitative assessment of these comparison, with details, we refer to~\cite{Aylott:2009ya}.


\subsection{Bayesian parameter estimation and model selection}

Bayesian inference is a powerful means of extracting information from observational data; although computationally too expensive for a search, it can be very useful for parameter estimation, in the study of candidates identified by search pipelines that are based on a pre-determined template bank.

Two different approaches were taken to make inferences on the NINJA data: (i) the estimation of the parameters of the signal assumed to be present in the data, and (ii) the calculation of confidence in the presence of the signal, quantified by the {\it odds ratio} between the signal and noise models of the data.

Both approaches require the calculation of the posterior probability-density function (PDF)  on the parameter space of the signal, given the data $d$:
\begin{equation}\label{pdf}
p(\vec{\theta}|d)=\frac{p(\vec{\theta})p(d|\vec{\theta})}{p(d)}  
\propto p(\vec{\theta}) 
\exp \left( -2 \int_0^\infty  \frac {\left|\tilde{d}(f)
    - \tilde{h}(f; \vec{\theta}) \right|^2}{S_\mathrm{n}(f)}\, df  \right)
\end{equation}
in the presence of Gaussian noise with power spectral density $S_\mathrm{n}(f)$, where $p(\vec{\theta})$ is the prior probability density of the parameters $\vec{\theta}$ and $h(\vec{\theta})$ is the model used to describe the signal.

A Markov-Chain Monte-Carlo (MCMC) method~\cite{vanderSluys:2007st,vanderSluys:2008qx} was used to coherently analyze data from multiple detectors and evaluate the posterior PDFs. 
This technique stochastically samples a 12-dimensional parameter space for the  best match to the data. The accuracy of the parameter estimation is assessed with a comparison of likelihoods at different points in parameter space.
Inspiral waveforms at 1.5 post-Newtonian order were used in the NINJA analysis, with good parameter estimation for injections with relatively low total mass, where the inspiral contained a significant fraction of the total signal-to-noise ratio.  For high-mass injections, matching the merger and ringdown portions of the waveform to inspiral templates resulted in poor parameter estimation.
More details on the MCMC code are available elsewhere in this volume~\cite{mcmcnrdaproc}.

Bayesian model selection, based on the nested sampling~\cite{Skilling:2004} Monte Carlo technique, was used to measure the confidence of a detection with a given template family.  
In this approach, the likelihood of signal and noise model are marginalized over a range of parameters, with their prior. 
%

The ratio of probabilities, or  \emph{posterior odds ratio}, of the two models  is the product of the \emph{prior odds ratio} and the \emph{Bayes factor}, computed from the data (eq.~\ref{pdf}) and corresponding to the level of confidence in the detection~\cite{Veitch:2008ur,Veitch:2008wd}.   
Bayes factors were computed for the injections in the NINJA data set with a model that included the coherent response of all four available interferometers and two non-spinning approximants: the standard (2~PN) TaylorF2 family, with inspiral truncated at $6M$, in $50 - 150 \Ms$, and the phenomenological inspiral-merger-ringdown templates described in~\cite{Ajith:2007kx} in $50 - 475 \Ms$, 
with uniform priors for all parameters.
%
For all the injections, the phenomenological approximants return a Bayes factor significantly larger than TaylorF2 approximants, which do not contain the merger and ring-down portion of the coalescence. 


The algorithm also yields the maximum likelihood estimate of the recovered parameters, with a measure of the statistical errors. 
The resulting total mass is in most cases underestimated,  but the sky location is on average fairly well determined, as it depends on the time of arrival of the signals at different instrument sites. This is currently under more careful investigation, for more detail please see, in this volume, ref.~\cite{AylottVeitchVecchio:2009}.

\section{Conclusions}\label{s:conclusions}
The NINJA project was conceived a first step towards a long term collaboration between numerical relativists and data analysts in an attempt to use numerical waveforms to enhance searches for gravitational waves and bridge the gap between gravitational wave observations and astrophysics.  
The work is unique in that it focused on running existing gravitational wave search algorithms on waveforms from numerical simulations.
To facilitate this first collaboration and encourage broad participation, the scope was limited, with a relatively small number of waveforms exchanged and limited coordination between groups. As a result, the statistics of simulations was too low to draw robust conclusions~\cite{Aylott:2009ya}.
NINJA has proven to be extremely valuable at framing the questions which need to be answered for an effective use of numerical relativity waveforms in data analysis: how many modes are needed for complicated waveforms? What accuracy is appropriate for each analysis? How important is hybridization with post-Newtonian waveforms? How can we best cover the parameter space? Which approximate waveform families are suitable for the
detection of {\it nature's} gravitational-wave signals, as modeled by
numerical-relativity waveforms, without significant loss of
signal-to-noise ratio? Which approximate families, if any, are
adequate for accurate parameter estimation?
Future NINJA analyses will build upon this work to address these questions.

\section*{Acknowledgements}

We thank the Kavli
Institute for Theoretical Physics (KITP) Santa Barbara for hospitality during
the workshop ``Interplay between Numerical Relativity and Data Analysis,''
where the NINJA project was initiated. The Kavli Institute is supported by
National Science Foundation grant PHY-0551164.

This project was supported in part by DFG grant SFB/Transregio~7
``Gravitational Wave Astronomy'' (BB, MH, SH, DP, LR, US); 
%
%
by National Science Foundation grants
PHY-0114375 (CGWP),
PHY-0205155 (UIUC),
PHY-0354842 (RM),
DMS-0553302 (MB, LB, TC, KM, HP, MS),
DMS-0553677 (LK, AM),
PHY-0553422 (NC),
PHY-0555436 (PL),
PHY-0600953 (PB, LG, RAM, RV),
PHY-0601459 (MB, LB, TC, KM, HP, MS),
PHY-0603762 (AB, EO, YP),
PHY-0649224 (BF),
PHY-0650377 (UIUC),
PHY-0652874 (FAU),
PHY-0652929 (LK, AM),
PHY-0652952 (LK, AM),
PHY-0652995 (MB, LB, TC, KM, HP, MS),
PHY-0653303 (PL, DS, MC, CL),
PHY-0653321 (VK, IM, VR, MvdS),
PHY-0653443 (DS),
PHY-0653550 (LC),
PHY-0701566 (ES),
PHY-0701817 (PB, LG, RAM, RV),
PHY-0714388 (MC, CL, YZ),
OCI-0721915 (ES),
PHY-0722315 (MC, CL),
PHY-0745779 (FP),
PHY-0801213 (FH),
PHY-0838740 (BF, LS, VR) 
and NSF-0847611 (DB, LP);
%
%
by NASA grants 
HST-AR-11763 (CL, MC, JF, YZ),
NNG-04GK54G (UIUC),
NNG-04GL37G (RM),
NNG-05GG51G (LK, AM),
NNG-05GG52G (MB, LB, TC, KM, HP, MS),
05-BEFS-05-0044 (GSFC),
06-BEFS06-19 (GSFC),
07-ATFP07-0158 (MC, CL, YZ),
and NNX-07AG96G (UIUC),
%
%
and by NSF cooperative agreement PHY-0107417 (DK, SC). 

%
%
BA was supported by a Vacation Bursary of the UK Engineering and Physical
Sciences Research Council. AV, JV and BS acknowledge support by the UK Science
and Technology Facilities Council.  SF acknowledges the support of the Royal
Society.  MH was supported by SFI grant 07/RFP/PHYF148.  FP 
acknowledges support from the Alfred P. Sloan Foundation. SH acknowledges
support from  DAAD grant D/07/13385, grant FPA-2007-60220 from the Spanish
Ministry of Science and Education and VESF.
MB, LB, TC, KM, HP and MS acknowledge support from the
Sherman Fairchild Foundation and the Brinson Foundation.  LK and AM
acknowledge support from the Fairchild Foundation.
BK was supported by the NASA Postdoctoral Program at the Oak Ridge Associated
Universities. SM was supported in part by the Leon A. Herreid Graduate 
Fellowship. RS was supported by an EGO sponsored fellowship, EGO-DIR-105-2007.

%
Computations were carried out under LRAC allocations 
MCA08X009 (PL, DS),
TG-MCA08X010 (FAU),
TG-MCA02N014 (LSU),
TG-MCA99S008 (UIUC),
TG-PHY990002 (MB, LB, TC, LK, KM, AM, HP, MS),
on LONI systems (LSU),
and on the clusters at the AEI, Cardiff
University, Northwestern University (NSF MRI grant PHY-0619274 to VK),
the LIGO Laboratory, 
NASA Advanced Supercomputing Division (Ames Research Center),
Syracuse University, 
LRZ Munich (BB, MH, SH, US), 
the University of Birmingham,
and the RIT NewHorizons cluster.

\section*{References}
\bibliography{bibliography/ninja}

\end{document}